# Continuous Real-Time Sensing with a Nitrogen Vacancy Center via Coherent Population Trapping

Shu-Hao Wu, Ethan Turner, and Hailin Wang
Department of Physics, University of Oregon, Eugene, Oregon 97403, USA

## Abstract

We propose and theoretically analyze the use of coherent population trapping of a single diamond nitrogen vacancy (NV) center for continuous real-time sensing. The formation of the dark state in coherent population trapping prevents optical emissions from the NV center. Fluctuating magnetic fields, however, can kick the NV center out of the dark state, leading to a sequence of single-photon emissions. A time series of the photon counts detected can be used for magnetic field estimations, even when the average photon count per update time interval is much smaller than 1. For a theoretical demonstration, the nuclear spin bath in a diamond lattice is used as a model fluctuating magnetic environment. For fluctuations with known statistical properties, such as an Ornstein-Uhlenbeck process, Bayesian inference-based estimators can lead to an estimation variance that approaches the classical Cramer-Rao lower bound and can provide dynamical information on a timescale that is comparable to the inverse of the average photon counting rate. Real-time sensing using coherent population trapping adds a new and powerful tool to the emerging technology of quantum sensing.

## I. INTRODUCTION

Quantum sensing exploits the sensitivity of a simple quantum system, such as solid-state spins, cold atoms, or superconducting circuits, to a given physical quantity to derive an estimate for the physical quantity. The near-term prospect for emerging quantum technologies and the potential for discoveries in wide ranging research areas have stimulated intense research efforts in quantum sensing [1,2]. One of the most promising systems for quantum sensing is a negatively charged nitrogen vacancy (NV) center in diamond, which can enable sensing of magnetic field, electric field, temperature, and strain with nanometer spatial resolution [3-6].

The most widely used quantum sensing approach has been Ramsey interferometry or Ramsey fringes, which probe the coherent time evolution of a single or a collection of two-level systems [3,7]. Incorporation of Bayesian phase estimations, adaptive measurements, and machine learning in Ramsey interferometry can further improve the sensitivity, increase the dynamical range, and reach the Heisenberg limited scaling of the underlying sensing process [8-14]. Ramsey interferometry is especially suitable for sensing of static as well as periodic signals. Repeated Ramsey interferometric measurements also allow the tracking of time-dependent fields [13,15,16]. Special Bayesian phase estimation and waveform reconstruction techniques have been developed for the time-dependent sensing [12,15,17]. Nevertheless, since each Ramsey interferometric measurement consists of three separate stages: initialization, coherent time evolution, and read-out, Ramsey interferometry cannot provide continuous real-time sensing.

Here we propose and theoretically analyze the use of coherent population trapping (CPT) in a NV center for continuous real-time sensing. In a CPT process, a $\Lambda$-type three-level system is trapped in a dark state, i.e., a special superposition of the two lower states, which is decoupled from the upper state due to destructive quantum interference [18]. The formation of the dark state prevents optical emissions from the NV center. A fluctuating magnetic environment, however, can kick the NV center out of the dark state, leading to a sequence of single-photon emissions from the NV center. These single-photon emissions carry the information on the fluctuating magnetic field and can in principle provide continuous real-time sensing of the magnetic field, as illustrated schematically in Fig. 1. Nevertheless, the overall collection/detection efficiency for optical emissions from a NV center is only a few percent under typical experimental conditions. For a realistic implementation, a primary challenge is to make continuous real-time estimations of the magnetic field with the few photons detected.

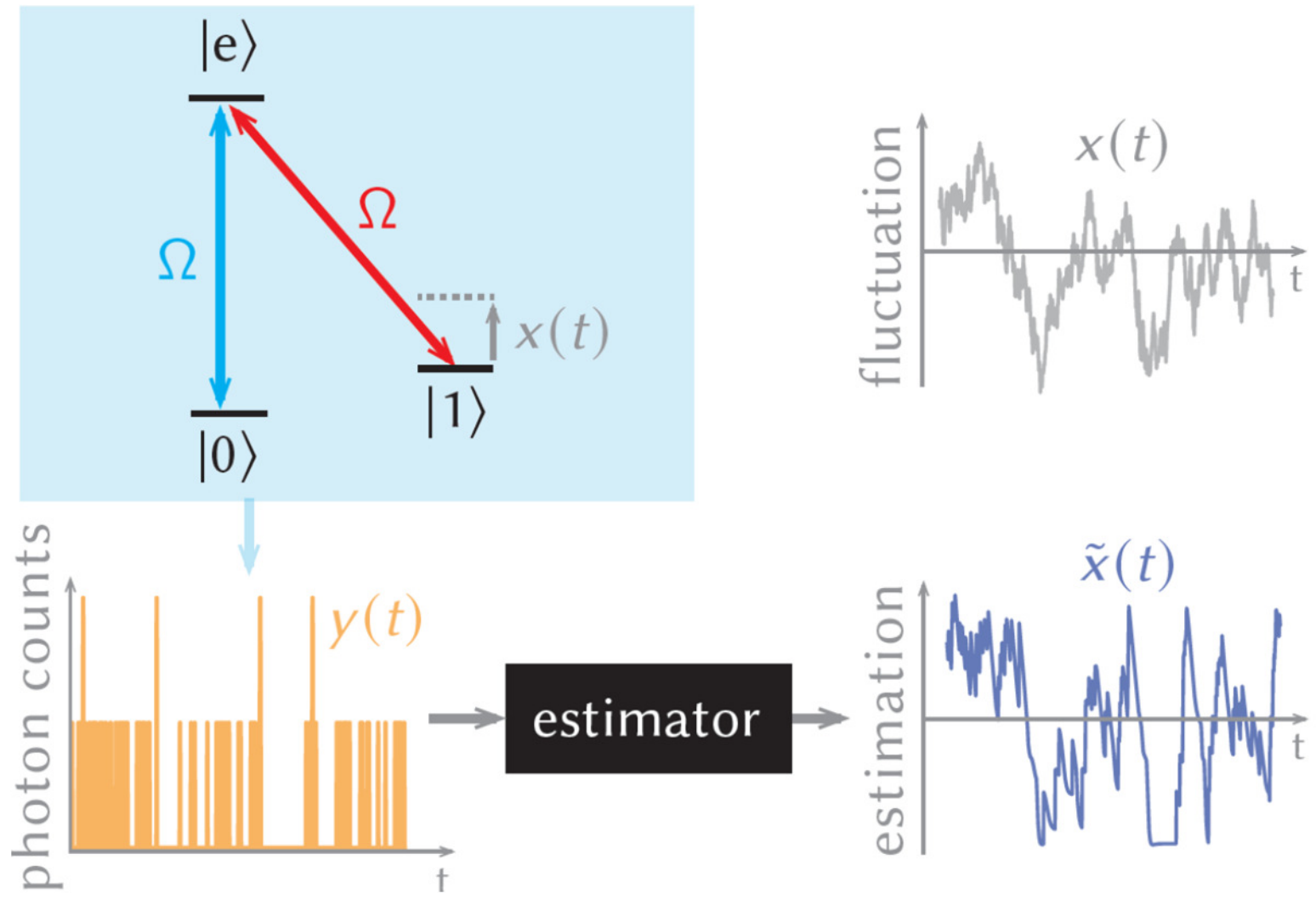

FIG. 1. Schematic of CPT-based continuous real-time sensing of the fluctuating magnetic environment at a NV center. The fluctuating magnetic field, which induces single-photon emissions from the NV center, can be estimated from the time series of the photon counts detected.

We show that this challenge can be overcome with the use of Bayesian inference. For the theoretical analysis, we have used the nuclear spin bath in diamond as a model fluctuating magnetic environment. We have employed the stochastic Schrodinger equation (SSE) to simulate single-photon emissions from a NV center in a CPT setting [19] and have treated the nuclear spin bath as an Ornstein-Uhlenbeck (OU) process [20-22]. Estimations of fluctuating magnetic fields are obtained from a time series of photon counts, for which the average number of photons per update time interval is much smaller than 1. By taking advantage of the known statistical properties of the fluctuating environment, we demonstrate that the Bayesian estimator can provide dynamical information on a timescale that is comparable to the inverse of the average photon counting rate. Additional theoretical analysis further indicates that the Bayesian estimator can approach the classical Cramer-Rao lower bound (CRLB).

Note that continuous sensing using resonant fluorescence of a two-level system has been proposed and analyzed in earlier studies, for which the distribution of waiting times between detected photon counts is used for Bayesian estimations of parameters such as Rabi frequency or laser detuning [23-26]. The CRLB is achieved in these estimations. This scheme has also been extended to a Λ-type three-level system, for which two-channel photon counting is used [25], though the role of CPT was not examined.

## II. PHYSICAL MODEL

### A) Coherent population trapping

For a NV center in diamond, several energy-level schemes have been used for the realization of CPT [27-31]. Without losing generality, here we consider two ground spin states, $m_s$=0 and $m_s$=1, coupling to an excited state, |$e$>, through two dipole optical transitions with frequencies, $\nu_0$ and $\nu_1$, respectively. Two external optical fields with frequencies, $\omega_0$ and $\omega_1$, couple to the two respective optical transitions with equal Rabi frequency Ω, as shown schematically in Fig. 1. With $\omega_0 \approx \nu_0$, $\omega_1 \approx \nu_1$, and $\rho_{ee} << 1$, the steady-state excited-state population is given by [30]

$$\rho_{ee} = \frac{\Omega^2}{2\Gamma\kappa}[1 - \frac{\Omega^2}{2\kappa}\frac{\gamma_s + \Omega^2/2\kappa}{(\delta - \omega_B)^2 + (\gamma_s + \Omega^2/2\kappa)^2}]. \tag{1}$$

where $\gamma_s$ and $\kappa$ are the decay rates for the spin coherence and optical dipole coherence, respectively, Γ is the spontaneous emission rate of the excited state, $\delta = \omega_0 - \omega_1$ is the laser detuning, and $\omega_B = \nu_0 - \nu_1$ is the frequency separation between the two spin states. CPT, which corresponds to the formation of a dark state for the two lower spin states, occurs near the Raman resonance with $\Delta = \delta - \omega_B = 0$. Note that the external optical fields lead to a power broadening described by the term $\Omega^2/2\kappa$ in Eq. (1). The CPT process can be characterized by a dimensionless cooperativity, defined as $C = \Omega^2/2\kappa\gamma_s$. CPT-based sensing will be carried out in the regime of $C$ >>1.

### B) Nuclear spin bath

We simulate the magnetic field fluctuations induced by the nuclear spin bath as an OU process characterized by a memory time $\tau_N$ [32], with

$$dx = -\frac{1}{\tau_N}xdt + \sqrt{\frac{2\sigma^2}{\tau_N}}dW_t, \tag{2}$$

where $x(t) = \omega_B(t) - \langle \omega_B(t) \rangle$ represents the bath-induced fluctuation in $\omega_B$, $dW_t$ denotes a Wiener increment that has a Gaussian distribution with mean zero and variance $dt$, and the autocorrelation function for $x(t)$ is characterized by

$$R(t) = \langle x(t_0)x(t_0 + t) \rangle = \sigma^2 e^{-|t|/\tau_N}. \tag{3}$$

For the two bath parameters, $\sigma$ can be derived directly from the decay of the Ramsey fringes and $\tau_N$ can be obtained from additional experiments of spin echoes [22]. For the numerical simulations in this paper, we take $\tau_N$ =1 ms and $\sigma/2\pi$= 0.13 MHz (which corresponds to a dephasing time, $T_2^*$ =1.7 μs [33]) for a $^{13}$C nuclear spin bath in diamond. Figure 2a shows an example of the simulated $x(t)$ and the autocorrelation function averaged over 1000 runs.

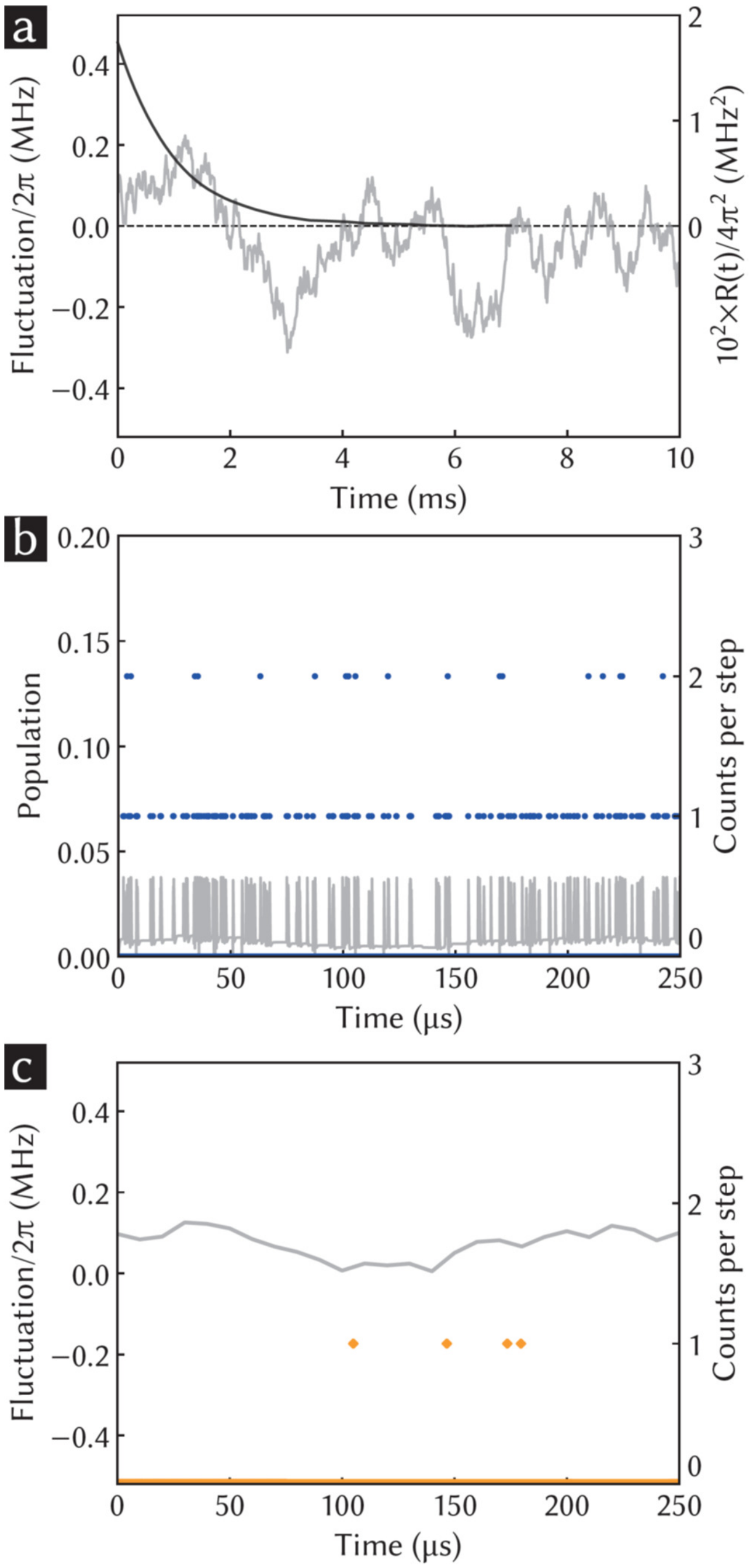

FIG. 2. (a) Simulated fluctuations in $\omega_B$ (grey line) and the auto-correlation $R(t)$ (black line). (b) The excited-state population (grey lines) as a function of time and the corresponding time series of photon counts emitted by the NV center (blue dots). (c) The time series of the photon counts detected (orange diamond) and the underlying fluctuation in $\omega_B$. The CPT parameters used are $(\Omega, \Gamma, \Delta_0)/2\pi$ =(2.8, 13, 0.25) MHz. The step size used for the photon counting is 0.1 μs.

**C) Time series of photon counts**

Magnetic field fluctuations induced by the nuclear spin bath of the diamond carbon lattice lead to corresponding fluctuations in $\omega_B$, which can kick a NV center out of or into the dark state in a CPT setting. This in turn induces fluctuations in the excited state population and leads to a sequence of single-photon emissions. Note that the nuclear spin bath fluctuates at a timescale many orders of magnitude slower than the NV radiative lifetime (12 ns [34]). It takes only a few spontaneous emission events for the NV center to reach the steady state, as confirmed in an earlier experimental study [35]. In this CPT setting, the single-photon emissions from the excited state carry the information on the fluctuating nuclear spin bath.

We have used the SSE to simulate single-photon emissions from a Λ-type three-level system in a CPT setting (see Appendix A), for which we take $\Gamma/2\pi$=13 MHz [34]. A Rabi frequency of $\Omega/2\pi$=2.8 MHz and a Raman detuning or bias of $\Delta_0 = \delta - <\omega_B> = 2\pi \cdot 0.25$ MHz are also used in the numerical simulation. Figure 2b shows the time series of the photons emitted by the three-level system, along with the corresponding excited state population. Only a small fraction of these photons is detected in a realistic sensing experiment. Figure 2c plots the time series of the photons detected, for which an overall collection/detection efficiency, $\eta$=1.6%, is used. Under these conditions, the average photon count rate is approximately 10000 per second. For comparison, the underlying fluctuations in $\omega_B$ are also plotted in Fig. 2c. Because of the very low photon counting rate, it is difficult to directly discern the pattern of the fluctuations from the time series of the detected photon counts.

Alternatively, we can simulate single-photon emissions using the population of the excited state with the assumption that the system has a Poisson distributed probability to emit a photon with the average detection rate given by $\eta\Gamma\rho_{ee}$. For this simulation, we make the approximation that $\rho_{ee}$ adiabatically follows the magnetic field of the bath, since the timescale of the bath fluctuation is orders of magnitude longer than the NV spontaneous emission lifetime. As shown in Appendix A, magnetic field estimations using the single-photon emissions obtained with the steady-state $\rho_{ee}$ exhibit the same behavior as those obtained with the SSE. For numerical calculations that require more than a few hundred runs, we have thus simulated single-photon emissions using the steady-state $\rho_{ee}$.

## III. ESTIMATORS

In this section, we discuss estimators, which we have used to output a time series of estimated frequencies, $\{\tilde{\mathbf{x}}_n\} = \{\tilde{x}_1, \tilde{x}_2, \ldots, \tilde{x}_n, \ldots\}$, from a time series of photon counts, $\{\mathbf{y}_n\} = \{y_1, y_2, \ldots, y_n, \ldots\}$, where $y_n$ is the number of photons detected during the $n$th time interval and the duration of the time interval $\tau$ is small compared with $\tau_N$. We will also discuss how close these estimations can be to the actual frequencies, $\{\mathbf{x}_n\} = \{x_1, x_2, \ldots, x_n, \ldots\}$.

### A) Bayesian estimator

We have used the Bayesian inference described by the Bayes update rule,

$$p(x_n \mid y_n, y_{n-1}, \ldots, y_1) \propto p_{\bar{y}_n}(y_n \mid x_n) \times p'(x_n \mid y_{n-1}, \ldots, y_1). \tag{4}$$

to estimate $x(t)$ from the time series of photon counts, where $p'(x_n \mid y_{n-1}, \ldots, y_1)$ is the prior probability distribution based on the previous time series of photon counts, $p(x_n \mid y_n, y_{n-1}, \ldots, y_1)$ is the posteriori probability distribution, and $p_{\bar{y}_n}(y_n \mid x_n)$ is the likelihood of detecting $y_n$ photons in the $n$th time interval given $x_n$. The likelihood follows a Poisson distribution,

$$p_{\bar{y}_n}(y_n \mid x_n) = \frac{\bar{y}_n^{\,y_n} e^{-\bar{y}_n}}{y_n!}, \tag{5}$$

where $\bar{y}_n = \eta\tau\Gamma\rho_{ee}(x_n)$ is the average number of detected photons expected for the given time interval and the update interval $\tau$ is sufficiently short such that $\bar{y}_n$ is small compared with 1. The estimation of $x(t)$ as a function of time is then given by

$$\tilde{x}(t) = \int p(x,t)x\,dx\,. \tag{6}$$

The prior probability distribution for $x_n$ can be improved if the statistical properties of the fluctuating field are known. For the OU process, the probability of finding $x_n$ at $t+\tau$ given $x_{n-1}$ at $t$ is given by a normal distribution,

$$p_{OU}(x_n, t+\tau \mid x_{n-1}, t) = N(x_{n-1}\,\mathrm{e}^{-\tau/\tau_N}, \sigma^2[1-\exp(-2\tau/\tau_N)]) \quad (7)$$

where the normal distribution has a mean of $x_{n-1}\,\mathrm{e}^{-\tau/\tau_N}$ and a variance of $\sigma^2[1-\exp(-2\tau/\tau_N)]$. In this case, the prior probability distribution for $x_n$ can be improved as

$$p'(x_n = x \mid y_{n-1},\ldots,y_1) = \int d\omega \cdot p(x_{n-1} = x-\omega \mid y_{n-1},\ldots,y_1)\cdot p_{OU}(x_n = x_{n-1}+\omega, t+\tau \mid x_{n-1}, t) \quad (8)$$

We will refer this improved Bayesian estimation process as the OU Bayesian estimator.

## B) Average count estimator

For comparison, we have also used the photon counts detected in a relatively long duration, $\tau_a$, to directly estimate $x(t)$. Specifically, we carry out the estimation by using

$$y_n^{(a)} = \eta\Gamma\tau_a\rho_{ee}(x_n), \quad (9)$$

where $y_n^{(a)}$ is the number of photons detected between time $n\tau-\tau_a$ and $n\tau$. For achieving acceptable photon counts (of order 10), we took $\tau_a = 100\tau$ for the results presented in section IV. This average count approach is not expected to be able to track bath fluctuations that occur in a relatively short timescale.

Note that electromagnetically induced transparency (EIT) of an ensemble of NV centers, which is closely related to CPT, has been used for static sensing [36]. CPT of an ensemble of NV centers, which can feature orders of magnitude greater photon counting rates than those of a single NV, can also be exploited for time-dependent sensing. A single quantum sensor, such as a single NV, however, is necessary for sensing microscopic fluctuations or for sensing at nanometer resolution.

## C) Cramer-Rao lower bound

How close an estimation is to the actual quantity is characterized by the average estimation variance, defined as $\mathrm{Var}[\tilde{x}_n(\mathbf{y})] = \langle(\tilde{x}_n(\mathbf{y}) - x_n(\mathbf{y}))^2\rangle$. Theoretically, the classical CRLB sets a lower bound on the variance. As shown in Appendix B, the CRLB for the OU-Bayesian estimation is given by

$$\mathrm{Var}[\tilde{x}_n(\mathbf{y})] \geq \frac{\sigma^2}{\sqrt{1+2\tau_N\eta\Gamma\sigma^2 g(\sigma)}}, \quad (10)$$

where $g(\sigma) = <(\partial\rho_{ee}/\partial x)^2/\rho_{ee}(x)>_x$.

For real-time sensing, only the historical data can be used. Data taken after the estimation event cannot be used for the estimation. However, the CRLB in Eq. (10) assumes that the entire data set can be used for the estimation. As shown in Appendix B, the CRLB when only the historical data can be used is revised as

$$\mathrm{Var}[\tilde{x}_n(\mathbf{y})] \geq \frac{\sigma^2}{\sqrt{1+2\tau_N \eta \Gamma \sigma^2 g(\sigma)}} \cdot \frac{2}{1+4/\sqrt{1+32\tau_N \eta \Gamma \sigma^2 g(\sigma)}}, \tag{11}$$

where we have assumed that $\tau_N \eta \Gamma \sigma^2 g(\sigma) > 2$ (which is satisfied in most cases). In the limit that $\tau_N \eta \Gamma \sigma^2 g(\sigma) >> 1$, the CRLB for estimations using the historical data set is twice that using the entire data set.

It should be noted that a necessary condition for achieving CRLB is that the posterior is a Gaussian distribution [37]. For our system, the posterior, $p(x_n | y_n, y_{n-1}, \ldots, y_1)$, is approximately Gaussian only when $\rho_{ee}$ depends linearly or quadratically on $x$.

## IV. RESULTS AND DISCUSSIONS

We have used the three estimators discussed in Section III for estimations of $\omega_B$ from a sequence of single-photon emissions. The estimations as a function of time shown in Fig. 3 are based on the simulated fluctuations and the corresponding time series of photon counts shown in Appendix A. The update time interval used is $\tau$=10 μs and the average photon count per time interval is 0.1. As shown in Fig. 3a, the estimations obtained with the average count estimator exhibit considerable deviations from the true frequency due to the low photon count rate. The estimations also feature significant delays with respect to the fluctuations in the actual frequency. The delays are due to the relatively long photon counting period, $\tau_a$, needed to avoid excessive fluctuations in the photon counts used for the estimation. In comparison, the estimations obtained with the simple Bayesian estimator provides a good estimate for the relatively longtime behavior of the frequency fluctuations. However, the photon counts are too low for the estimator to provide dynamical information on a relatively short timescale.

Much improved estimations are obtained with the OU Bayesian estimator, as shown in Fig. 3b. By taking advantage of the known statistical properties of the OU process, the Bayesian estimator is able to yield dynamical information on a timescale much shorter than that is achieved with the simple Bayesian estimator. The OU Bayesian estimator provides continuous real-time sensing of the fluctuating magnetic field, even when the timescale of the fluctuation is comparable to the inverse of the average photon count rate. It should be noted that an earlier study has used the complete CPT spectrum of a single NV for the sensing of the nuclear spin bath in diamond [28]. As shown in [28], it takes about 5 ms and about 100 detected photons to get a single CPT spectrum and thus to obtain a single estimation.

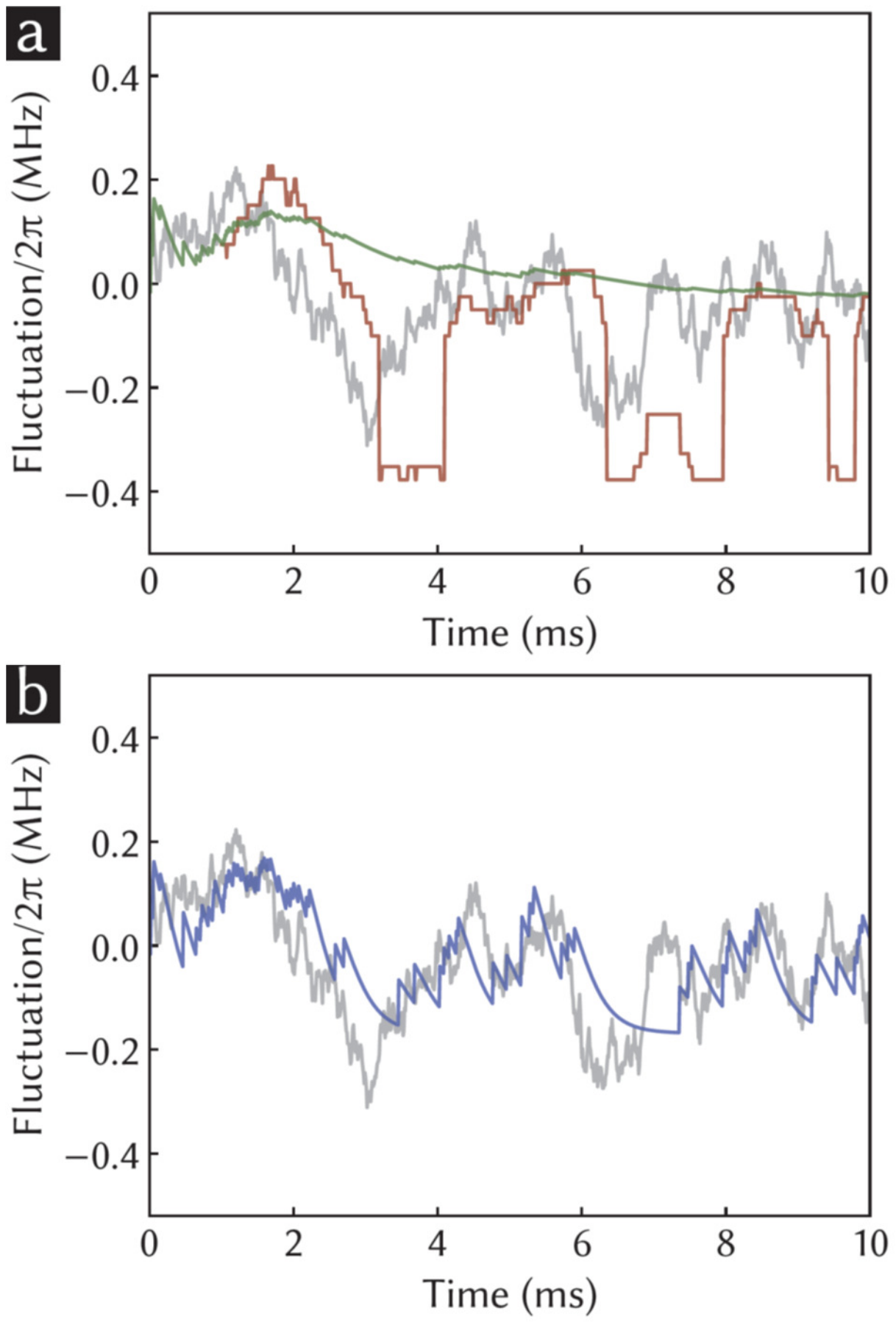

FIG. 3. (a) Estimations obtained with the average photon account estimator (brown line) and the simple Bayesian estimator (green line). (b) Estimations obtained with the OU Bayesian estimator (blue line). For comparison, the grey line plots the actual fluctuations. The CPT and bath parameters used are the same as those for Fig. 2.

For a quantitative analysis of the effectiveness of the estimators, we have numerically calculated the estimation variances and have compared these variances with those expected from the CRLB. The variances shown in Fig. 4 are the averaged results of 100 runs. Each run covers the time duration from $t$=2 ms to $t$=10 ms and uses sequences of single photon emissions generated from the steady-state $\rho_{ee}$, as discussed in Section II.C. The parameters used for the bath and the CPT process are the same as those used for Fig. 2 unless otherwise specified.

Figure 4a shows the variances obtained with the three different estimators as a function of the memory time, $\tau_N$, of the nuclear spin bath. For comparison, Fig. 4a also plots the theoretically expected CRLB for the OU Bayesian estimator and the variance, $\sigma^2$, of the actual frequency fluctuations. As expected, the variances obtained with the average count estimator are far above the CRLB. These variances also exceed $\sigma^2$. The variances obtained with the simple Bayesian estimator fall below $\sigma^2$ but are still far above the CRLB. In comparison, the variances obtained with the OU Bayesian estimator nearly approach the CRLB. Both the variances and the CRLB improve gradually with increasing $\tau_N$. By taking advantage of the known statistical properties of the fluctuating fields, the OU Bayesian estimator extracts nearly the maximum amount of information from the detected single photons.

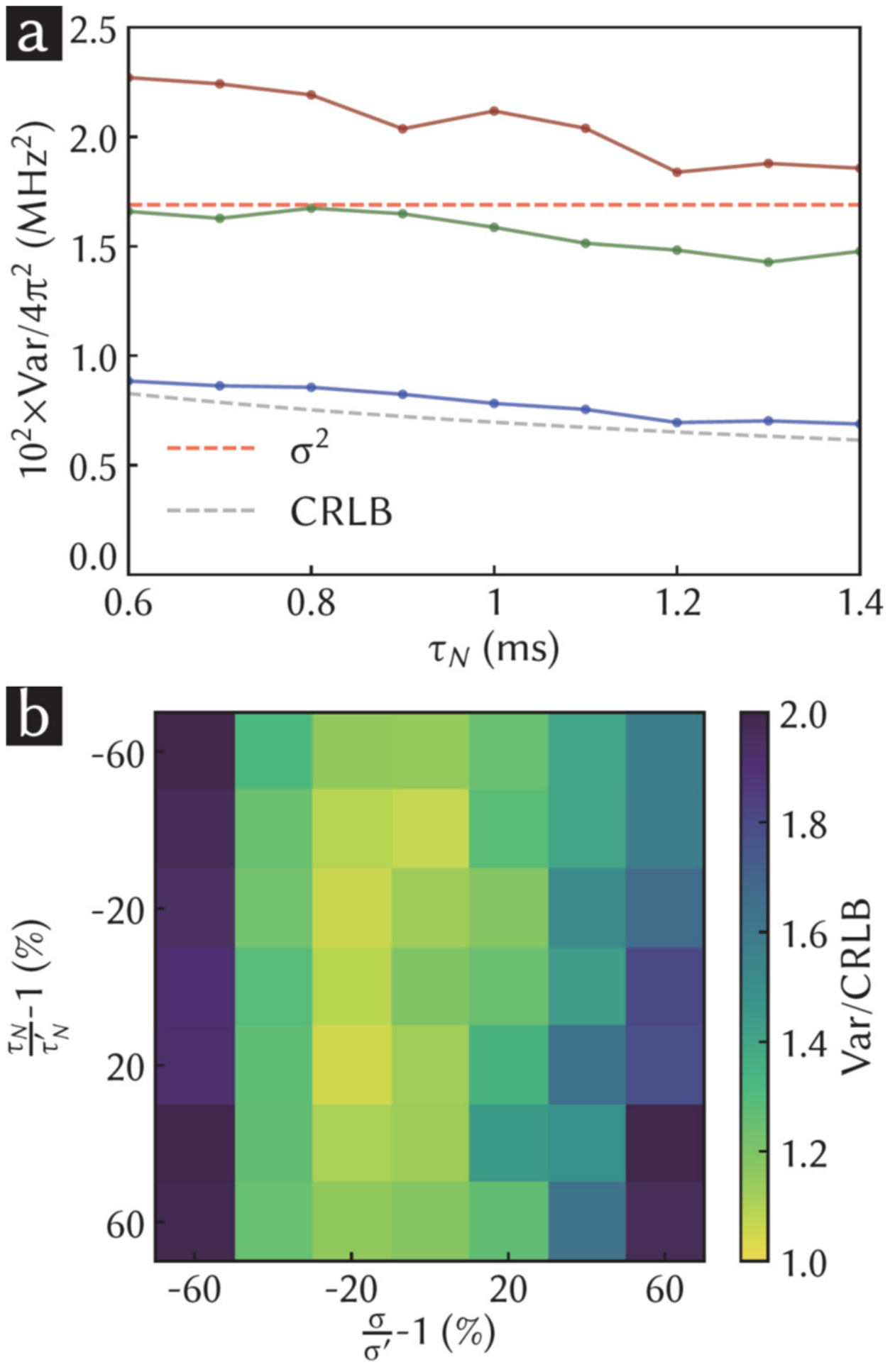

FIG. 4. (a) The average variances for estimations obtained with the average count estimator (brown dots), simple Bayesian estimator (green dots), and OU Bayesian estimator (blue dots) as a function of the bath memory time $\tau_N$. The estimation variances are also compared with the CRLB of the OU Bayesian estimator and with $\sigma^2$, the variance of the fluctuating $\omega_B$. (b) The ratio of the estimation variance over the CRLB as $\tau_N$ and $\sigma$ used in the OU Bayesian estimator, with $\tau_N$= 1 ms and $\sigma/2\pi$ =0.13 MHz, deviate from the true parameters, $\tau'_N$ and $\sigma'$.

For the OU Bayesian estimator, the estimations are expected to depend on deviations of the statistical parameters used in the estimator from the true parameters. Figure 4b shows the ratio of the estimation variances over the corresponding CRLB as $\tau_N$ and $\sigma$ used in the estimator move away from the true parameters. Variances close to the CRLB can still be obtained when the deviations of $\tau_N$ and $\sigma$ from the true parameters are within 20%. The estimation variances, however, can exceed those obtained from the simple Bayesian estimator when $\tau_N$ and $\sigma$ deviate from the true parameters by more than 50%. As discussed earlier, $\tau_N$ and $\sigma$ can be deduced from experiments such as Ramsey fringes and spin echoes. For the experimental implementation of the OU Bayesian estimator, feedback control, which uses the estimations to keep the NV in the dark state, can also be employed for the further optimization of the statistical parameters used in the estimator. In addition, the feedback control can also serve as a verification for the estimations.

The choice of the CPT parameters, including Rabi frequency $\Omega$ and bias $\Delta_0 = \delta - <\omega_B>$, also strongly affects the effectiveness of the estimations. For example, little information on the magnetic field fluctuations can be obtained from a sequence of single-photon emissions when $\Delta_0$ is set to near 0. Figure 5a shows the variances for the OU Bayesian estimator as a function of $\Omega$ and $\Delta_0$, with all other parameters the same as those in Fig. 2. In this case, the optimal estimation is achieved with $\Omega/2\pi$= 2.5 MHz and $\Delta_0/2\pi$= 0.2 MHz. In comparison, the corresponding CRLBs shown in Fig. 5b decreases with decreasing $\Omega$ and $\Delta_0$, which is expected from Eq. (11). As can be seen from Eq. (11), the CRLB is minimized when $g(\sigma) = <(\partial\rho_{ee}/\partial x)^2/\rho_{ee}(x)>_x$ reaches maximum.

The optimal CPT parameters for the OU Bayesian estimator shown in Fig. 5a reflect a tradeoff between a steep $\partial\rho_{ee}/\partial x$ and the need to avoid the zero bias region. As discussed in more detail in Appendix C, this also corresponds to the tradeoff between the working range of the sensing process and how sensitive the sensing process can be. Note that as discussed in Section

III.C, the CRLB is only achievable when the posterior is a Gaussian distribution [37], which occurs when $\rho_{ee}$ depends either linearly or quadratically on $x$. Near zero bias, the CRLB can no longer been achieved.

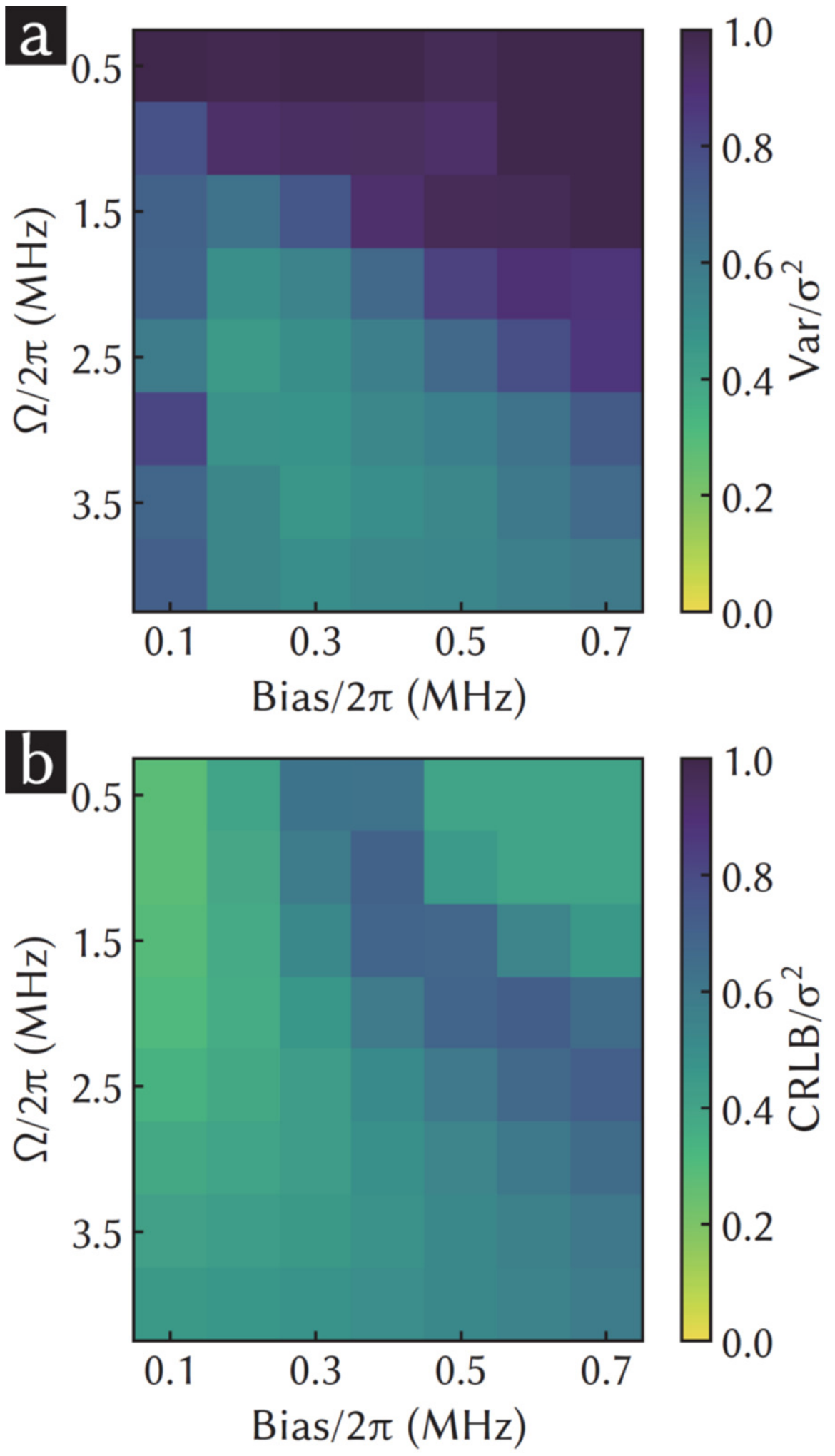

FIG. 5. (a) Estimation variances vs Ω and bias (i.e., $\Delta_0$). The color bar corresponds to Var/$\sigma^2$. (b) CRLBs for the OU Bayesian estimator vs Ω and bias. The color bar corresponds to CRLB/$\sigma^2$.

## IV. CONCLUSION

We have shown theoretically that CPT can enable continuous real-time sensing using a single quantum sensor such as a NV center in diamond. The use of Bayesian inference for the real-time sensing allows effective estimation of magnetic fields from a time series of detected photon counts, for which the average photon count per time interval is much smaller than 1, and can lead to estimation variances that approach the CRLB and to a time resolution that is comparable to the inverse of the average photon counting rate. While a NV center along with the surrounding nuclear spin bath has been used as a model system for the theoretical analysis, other spin systems such as defect centers in SiC can also serve as CPT-based quantum sensors [38]. In addition to the nuclear spin bath, the real-time sensing can be applied to studies of time-varying magnetic field changes or fluctuations in a variety of systems, for example, two-dimensional materials on a diamond surface [39,40]. Real-time sensing of a single nuclear spin is also possible, provided that the nuclear spin in question features a magnetic field that exceeds considerably that induced by the nuclear spin bath [41,42]. Together with feedback control techniques, the real-time sensing can also be used for the protection of a spin qubit from the surrounding magnetic fluctuations [16]. Overall, CPT-based real-time sensing can add a new and powerful tool to the emerging technology of quantum sensing.

## ACKNOWLEDGEMENTS

We acknowledge helpful discussions with Jonathan Dowling, Arshag Danageozian, and Pfaff Wolfgang. This work is supported by a grant from ARO MURI.

## APPENDIX A: Simulation of time series of photon counts

For the Λ-type three-level system in Fig. 1 and with $\omega_0 = \nu_0$, the Hamiltonian in the rotating frame is given by

$$H(t) = \frac{\hbar\Omega}{2}(|e\rangle\langle 0| + |e\rangle\langle 1| + |0\rangle\langle e| + |1\rangle\langle e|) + \hbar[\Delta_0 + x(t)]|1\rangle\langle 1|. \tag{A1}$$

For CPT-based real-time sensing, we operate the CPT process with $C$>>1. In this regime, the power broadening far exceeds the intrinsic decoherence rate ($\gamma_s / 2\pi \approx 0.1$ kHz [43]). We will thus set $\gamma_s$=0 in our simulation. The master equation is then given by

$$\dot{\rho}(t) = -\frac{i}{\hbar}[H(t), \rho(t)] + \frac{\Gamma}{2} D(|0\rangle\langle e|)\rho(t) + \frac{\Gamma}{2} D(|1\rangle\langle e|)\rho(t), \tag{A2}$$

where $D(\hat{O})\rho(t) = \hat{O}\rho(t)\hat{O}^{+} - \{\hat{O}^{+}\hat{O}, \rho(t)\}/2$ and we have also assumed $\kappa = \Gamma/2$.

We have used the SSE to simulate single-photon emissions from the three-level system. The SSE, which tracks every collapse of the system, is given by (with $\hbar = 1$)

$$d|\psi\rangle = -i\left(H - \frac{i\Gamma}{2}|e\rangle\langle e|\right)|\psi\rangle dt + \frac{\Gamma}{2}\langle |e\rangle\langle e|\rangle |\psi\rangle dt$$
$$+\left(\frac{|0\rangle\langle e|}{\sqrt{\langle |e\rangle\langle e|\rangle}} - 1\right)|\psi\rangle dN_0 + \left(\frac{|1\rangle\langle e|}{\sqrt{\langle |e\rangle\langle e|\rangle}} - 1\right)|\psi\rangle dN_1, \tag{A3}$$

where $N_0$ and $N_1$ are the accumulated photon counts from the $|e\rangle \leftrightarrow |0\rangle$ and $|e\rangle \leftrightarrow |1\rangle$ transitions, respectively. Without any optical emission, the system evolves under an effective Hamiltonian, $H - i\Gamma(|e\rangle\langle e| - \langle |e\rangle\langle e|\rangle)/2$. When a photon is emitted via the $|e\rangle \leftrightarrow |0\rangle$ (or $|e\rangle \leftrightarrow |1\rangle$) transition, the system collapse to the $|0\rangle$ (or $|1\rangle$) state.

Figure 6a shows, along with the underlying fluctuations in $\omega_B$, the calculated time series of photons detected with $\eta$=1.6% and with the collection and detection loss modeled as a random process. The parameters used are the same as those for Fig. 2c, except that the step size used for the photon counting is 10 μs for Fig. 6a and is 0.1 μs for Fig. 2c. The data set in Fig. 6a is used for the estimations shown in Fig. 3.

We have also used the steady-state population of the excited state to model the single photon emission process. In this case, the steady-state solution to Eq. (A2) is obtain for a given $x_n$. The Λ-type three-level system in the excited state has a Poisson distributed probability to emit photons, with a mean detected photon count, $\bar{y}_n = \eta\tau\Gamma\rho_{ee}(x_n)$, for a given time interval $\tau$. Figure 6b compares the estimation variances derived from the time series of photon counts obtained with the SSE and those obtained with the steady-state population, for which we have used the OU Bayesian estimator and the same parameters as those for Fig. 6a. As can be seen from Fig. 6b, estimations obtained with these two simulation approaches exhibit essentially the same behaviors.

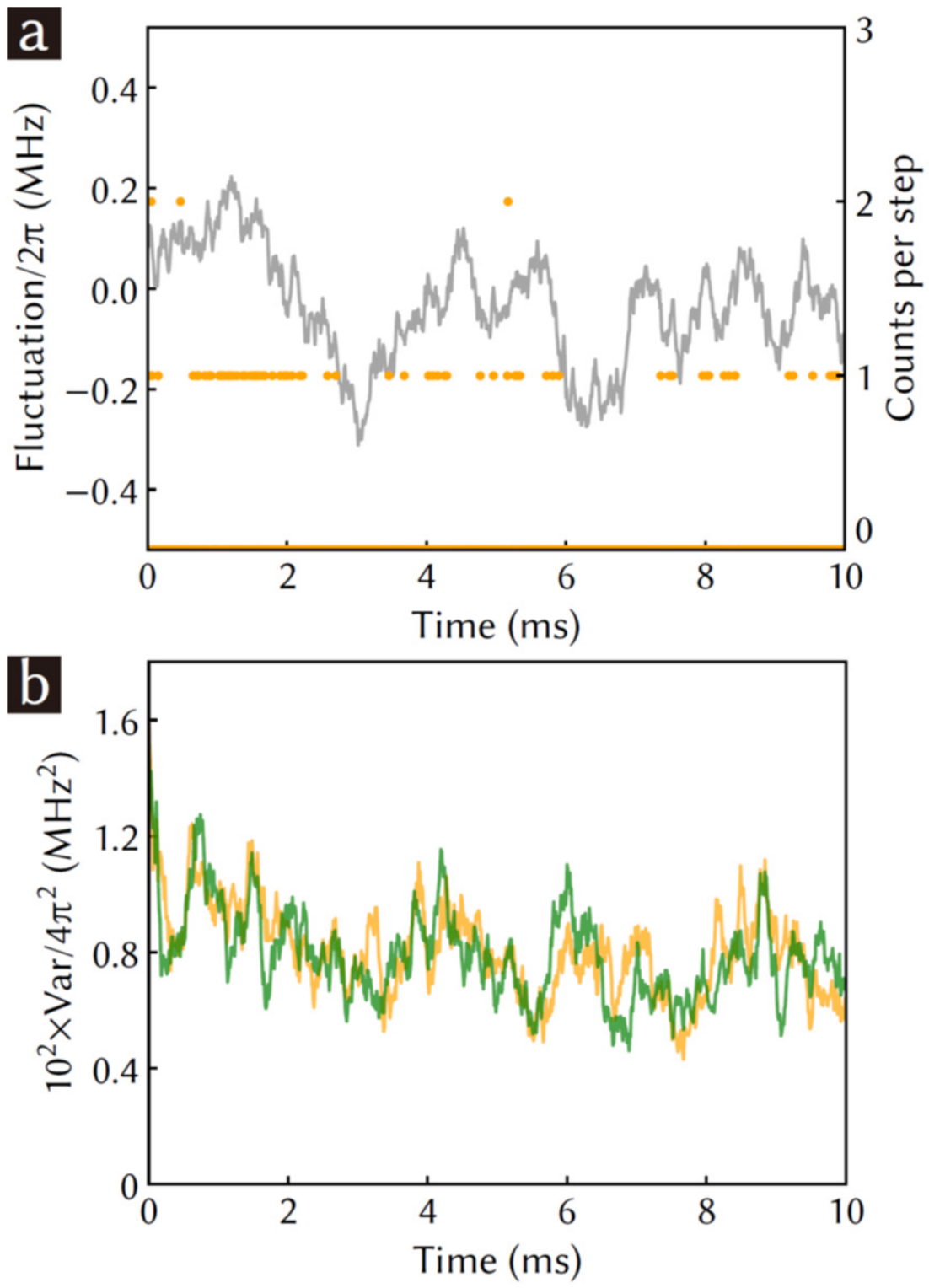

FIG. 6. (a) Time series of photon counts generated by the SSE (orange dots), along with the underlying fluctuations in $\omega_B$ (grey lines). The step size for photon counting is 10 μs. (b) Estimation variances obtained from time series of photon counts generated by the SSE (orange lines) and those by the steady-state population of the excited state (green lines). The results shown are averages over 100 runs. All parameters used are the same as those for Fig. 3.

## APPENDIX B: Derivation of the Cramer-Rao Lower Bound

For the estimation of a time series of parameters, there exists a theoretical lower bound for the estimation variance (i.e. the CRLB), $\mathrm{Var}[\tilde{x}_i(\mathbf{y})] \geq (\mathbf{F}^{-1})_{ii}$, where $(\mathbf{F}^{-1})_{ii}$ is the diagonal component of the inverse Fisher information matrix. The Fisher matrix consists of two parts, $\mathbf{F}_{ij} = \mathbf{F}_{ij}^M + \mathbf{F}_{ij}^B$, with

$$\mathbf{F}_{ij}^M = \langle \frac{\partial \ln p(\mathbf{y} \mid \mathbf{x})}{\partial x_i} \frac{\partial \ln p(\mathbf{y} \mid \mathbf{x})}{\partial x_j} \rangle, \tag{B1}$$

$$\mathbf{F}_{ij}^B = \langle \frac{\partial \ln p(\mathbf{x})}{\partial x_i} \frac{\partial \ln p(\mathbf{x})}{\partial x_j} \rangle. \tag{B2}$$

For the OU-Bayesian estimations, $\mathbf{F}_{ij}^M$ represents the information provided by the time series of the photon counts, whereas $\mathbf{F}_{ij}^B$ contains the information provided by the correlations inherent in the fluctuations [37].

We can evaluate the Fisher matrix analytically, with $\mathbf{F}_{ij}^M = \Gamma \eta \tau g(\sigma) \delta_{ij}$ [44]. In the limit that the update interval $\tau \to 0$, we have

$$F^M(t_i, t_j) = \lim_{\tau \to 0} \frac{\mathbf{F}_{ij}^M}{\tau^2} = \Gamma \eta g(\sigma) \delta(t_i - t_j). \tag{B3}$$

In this limit, the Fourier transform of $\mathbf{F}_{ij}^B$ is given by [44],

$$\mathcal{F}\left\{F^B(t_i, t_j)\right\} = \mathcal{F}\left\{\lim_{\tau \to 0} \frac{\mathbf{F}_{ij}^B}{\tau^2}\right\} = \frac{1}{2\pi \mathcal{F}\left\{\sigma^2 e^{-|t_i - t_j|/\tau_N}\right\}}. \tag{B4}$$

Using the above results, we have derived the lower bound for the estimation variance,

$$\mathrm{Var}[\tilde{x}_t(\mathbf{y})] \geq F^{-1}(0) = \frac{\sigma^2}{\sqrt{1 + 2\Gamma \eta \sigma^2 \tau_N g(\sigma)}}. \tag{B5}$$

The CRLB in Eq. (B5) is for the entire data set and occurs near the middle of the time series. However, only the historical information can be used in real-time sensing. In this case, the estimation effectively takes place at the end of the available data set. The CRLB for the real-time sensing can thus be obtained via the ratio between $\mathbf{F}_{T,T}^{-1}$, the diagonal element of the inverse Fisher matrix at the end of the data set, and $\mathbf{F}_{T/2,T/2}^{-1}$, the diagonal element of the inverse Fisher matrix at the middle of the data set.

It can be shown that $\mathbf{F}_{T,T}^{-1} / \mathbf{F}_{T/2,T/2}^{-1} = \det(\mathbf{F}'_T) / \det(\mathbf{F}'_{T/2})^2$, where

$$\mathbf{F}'_T = \begin{pmatrix} 1-\tau_0^4 & \tau_0^3-\tau_0 & & & \\ \tau_0^3-\tau_0 & 1-\tau_0^4 & \ddots & & \\ & \ddots & \ddots & \ddots & \\ & & \ddots & 1-\tau_0^4 & \tau_0^3-\tau_0 \\ & & & \tau_0^3-\tau_0 & 1-\tau_0^2 \end{pmatrix} + g_0 \begin{pmatrix} 1 & & & \\ & \ddots & & \\ & & \ddots & \\ & & & \ddots & \\ & & & & 1 \end{pmatrix} \quad \text{(B6)}$$

is a matrix of order $n = T/\tau$, with $g_0 = \Gamma\eta g(\sigma)\tau\sigma_0^4/\sigma^2$, $\tau_0 = e^{-\tau/\tau_N}$, and $\sigma_0^2 = \sigma^2(1-\tau_0^2)$. Using recurrence relations, we obtain

$$\frac{\mathbf{F}_{T,T}^{-1}}{\mathbf{F}_{T/2,T/2}^{-1}} = \frac{(Ar_1^n + Br_2^n)}{(Ar_1^{n/2} + Br_2^{n/2})^2}, \quad \text{(B7)}$$

where $A$, $B$, $r_1$, $r_2$, which are determined by the boundary conditions, are functions of $\tau_0$, $g_0$. In the limit that $n$>>1, we have

$$\frac{\mathbf{F}_{T,T}^{-1}}{\mathbf{F}_{T/2,T/2}^{-1}} \approx \frac{Ae^{TC_+} + Be^{TC_-}}{A^2e^{TC_+} + B^2e^{TC_-} + 2ABe^{T(C_+ + C_-)/2}}, \quad \text{(B8)}$$

where $C_\pm = (-8 \pm \sqrt{1+32\Gamma\eta g(\sigma)\sigma^2\tau_N})/4\tau_N$. For our choice of parameters, $\Gamma\eta g(\sigma)\sigma^2\tau_N > 2$. The CRLB when only historical information can be used is then given by

$$\mathbf{F}_{T,T}^{-1} \geq F^{-1}(0) \cdot \lim_{T\to\infty} \frac{Ae^{TC_+} + Be^{TC_-}}{A^2e^{TC_+} + B^2e^{TC_-} + 2ABe^{T(C_+ + C_-)/2}} = \frac{2F^{-1}(0)}{1 + 4/\sqrt{1+32\Gamma\eta\sigma^2\tau_N g(\sigma)}} \quad \text{(B9)}$$

## APPENDIX C: Tradeoff between working range and sensitivity

For sensing of a constant parameter, the minimum detectable signal, which scales inversely with the total detection time, is defined as the signal amplitude that yields unit signal-to-noise ratio and the sensitivity is defined as the minimum detectable signal per unit time [1]. However, this definition of sensitivity is no longer applicable for sensing of a time-varying parameter with a given distribution, for which we would like to know the estimation performance averaged over the entire distribution rather than that at a single point. In this case, we can choose the estimation variance $\mathrm{Var}[\tilde{x}_n(\mathbf{Y})]$ as an effective measure of how good the sensing process is [17]. Note that $\sqrt{\mathrm{Var}[\tilde{x}_n(\mathbf{Y})]}$ is not the minimum detectable signal.

Although it is difficult to define a sensitivity for the sensing of a time-varying parameter, for our case it is apparent that the sharper the CPT dip is (more precisely the greater $(\partial\rho_{ee} / \partial x)^2 / \rho_{ee}(x)$ is), the more sensitive the sensing process becomes. In comparison, the working range of the sensing process, which is the maximum range of the distinguishable signal, scales with the width of the CPT dip. There is thus a trade-off between the working range and how sensitive the sensing process is. Since the power broadening of the CPT dip is given by $\Omega^2 / \Gamma$, the optimal choice of Ω for the sensing process reflects this tradeoff.

To illustrate this tradeoff, Fig. 7 shows the calculated estimation variance as a function of Ω, for which the optimal bias, $\Delta_0$, is used for each Ω. When Ω is too small, the working range is small compared with the actual range of the frequency fluctuation, which is a few times of σ. Whereas when Ω is too large, the sensing process becomes relatively insensitive. Both cases result in relatively large $\mathrm{Var}[\tilde{x}_n(\mathbf{Y})]$, as shown in Fig. 7.

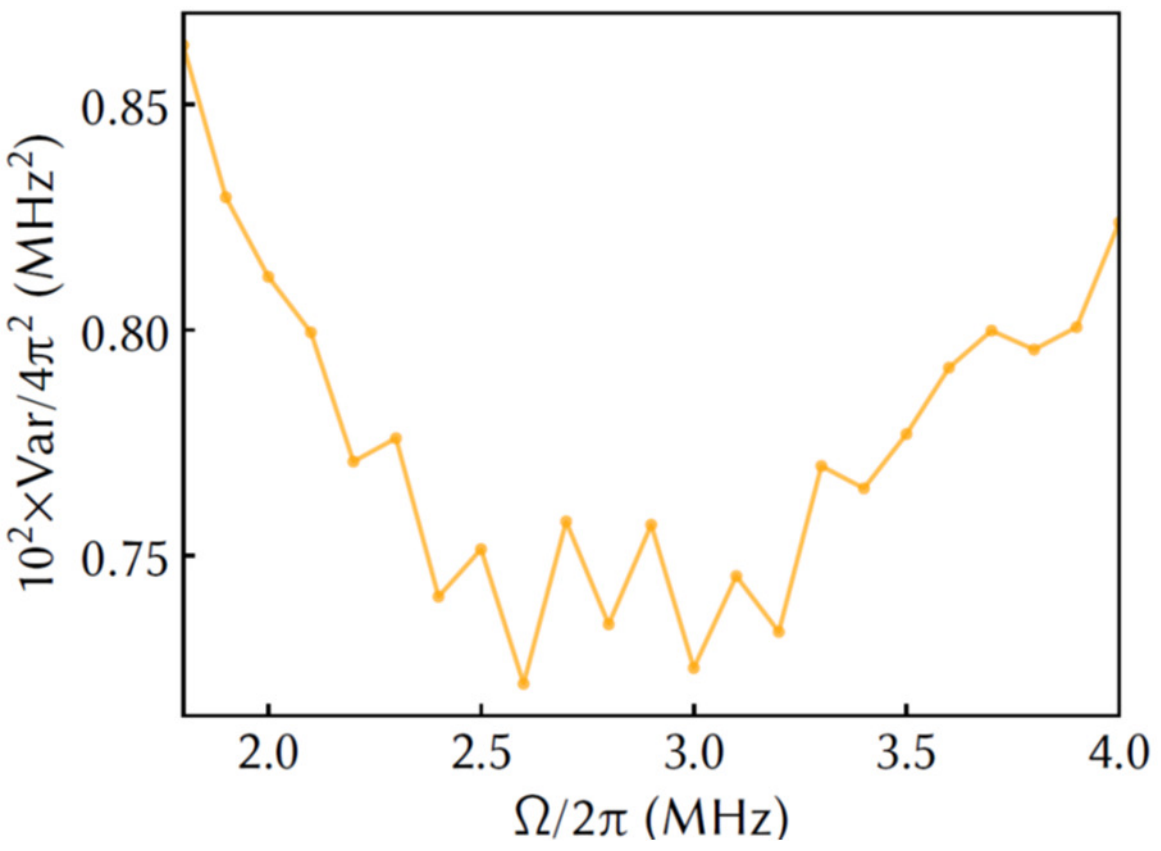

FIG. 7. Estimation variances vs Ω, for which the optimal $\Delta_0$ is used for each Ω, with other parameters the same as those used for Fig. 5a.